\documentstyle[preprint,aps]{revtex}

\def\tj#1#2#3#4#5#6{\left(\matrix{{#1} & {#2} & {#3} \cr {#4} & {#5} &
                                      {#6} \cr}\right)}
                                      \def\bqn{\begin{equation}}
                                      \def\nqn{\end{equation}}
                                      \def\bar#1{\overline{#1}}
\def\vec#1{{\bf#1}}
\def\vO{\vec \Omega}

\begin{document}
\draft \preprint{} \title{Response Function of the
Fractional Quantized
Hall State on a Sphere II: Exact Diagonalization} \author{Song~He}
\address{ AT\&T Bell Laboratories, Murray Hill, NJ 07974 \\ }
\author{S.~H.~Simon and B.~I.~Halperin} \address{ Department of
Physics, Harvard University \\ Cambridge, MA 02138 } \date{Feb 21,
1994} \maketitle

\begin{abstract}
We study the excitation spectra and the dynamical structure factor of
quantum Hall states in a finite size system through exact
diagonalization. Comparison is made between the numerical results so
obtained and the analytic results obtained from a modified RPA in the
preceding companion paper\cite{Simon}. We find good
agreement between the results at low energies.
\end{abstract}

\narrowtext
\section{Introduction}
\label{Intro}

By using a singular gauge transformation to convert electrons into
fermions interacting through a Chern-Simons field, much progress has
been made in understanding the physics of a two-dimensional
interacting electron systems in strong magnetic fields at simple
filling fractions\cite{Simon,HLR,Simon2}. At the mean field level,
this transformation maps certain fractional quantized Hall states into
integer quantized Hall states of transformed fermions so that some
conventional many-body techniques may be used to tackle the problem.
In the preceding companion paper\cite{Simon}, an analytic procedure
call the modified random phase approximation (modified RPA) has been
constructed in the spherical geometry for evaluating the response of
certain quantized Hall states to external perturbations. The purpose of
this paper is to present numerical results of excitation spectra and
the dynamical structure factors at selected filling fractions, calculated
using exact diagonalization and the modified RPA, so that comparison
can be made and the performance of the modified RPA can be evaluated.
Specifically, we present numerical results from both exact
diagonalization and modified RPA for systems of $N=8$ and $N=12$
electrons at filling factor $\nu=1/3$ and $\nu=3/7$ respectively.

As usual for numerical calculations, we work in the limit where the
electron-electron interaction is weak compared to the cyclotron
energy.  For the exact diagonalization method, this means that all
electrons are confined to the lowest Landau level.  Within the
modified RPA calculations, this means that we are working in the limit
where the bare band mass $m_b$ is infinitely small compared to the
effective mass $m^*$, whose scale is set by the electron-electron
interaction strength.  (In practice, we have carried out these
calculations for the value $m^*/m_b = 50$.)

Our main purpose is to compare the exact and approximate calculations
of the dynamical structure factor $S(q,\omega)$ over a range of
wavevectors, for the frequencies which are on the scale of the
electron-electron energy, and therefore very much lower than the
cyclotron frequency $\omega_c$.  For such frequencies, the structure
factor $S(q,\omega)$ is equivalent to the projected structure factor
$\bar S(q,\omega)$, in the exact calculation where all electron
operators are projected onto the lowest Landau level.  To obtain
additional information, however, we wish to compare various frequency
moments of the projected structure factor with the corresponding
quantities obtained from the modified RPA.  (Specifically, we
calculate contributions to the $f$-sum rule, the projected static
structure factor $\bar S(q)$ , and the static wavevector-dependent
compressibility.)  For these comparisons, we define the projected
structure factor in the modified RPA by excluding the contribution of
the ``Kohn mode'', which occurs at the cyclotron frequency at $q
\approx 0$.

Our plan for this paper is as follows. In section \ref{Model}, we
describe the model and briefly discuss the method we use in the exact
diagonalization.  In addition, we review a few necessary points of the
modified RPA calculation.  In section \ref{Results}, we present
results for the projected dynamical structure factor $\bar{S}(q,\omega)$
which describes the low energy response of the system to an external
perturbation and gives the dispersion of the collective modes of the
system. As well as showing $\bar{S}(q,\omega)$, we will also calculate
the dispersion of the collective mode in the single mode approximation
(SMA), the contributions of the excitations in the lowest Landau level
to the projected static structure factor $\bar{S}(q)$, the $f$-sum
rule, and the compressibility sum rule.  Our conclusions are
summarized in section \ref{C}.

\section{Model and method}
\label{Model}

We will use the spherical geometry throughout this paper, so our
system is given by a magnetic monopole of total flux $N_{\phi}$
quanta at the center of the sphere with $N$ electrons confined to the
sphere's surface, interacting via a Coulomb potential.  If the system
were infinitely large (ie, a planar system), the filling fraction
would be given by the ratio of the number of electrons to the number
of flux quanta ($\nu =\frac{N}{N_{\phi}}$).  However, in a finite
sized system, this relation is changed to\cite{Haldane,Fano,Morf}
\begin{equation}
  \label{eq:Nshift1}
  N_{\phi}  = [\nu]^{-1} N - X(\nu).
\end{equation}
where $X(\nu)$ is known as the ``shift'' \cite{Wen} of the state, and
depends also on the topology of the system.  In this work, we will be
interested in filling fractions of the form $\nu_p = \frac{p}{2p+1}$.
For these states, in our spherical geometry, the shift is given by
\begin{equation}
  \label{eq:Nshift2}
  X(\nu _p) = 2 + p.
\end{equation}

In the spherical geometry, by rotational invariance, the eigenstates
of the system can be classified by the conserved quantum numbers
$(L,M)$, where $L$ is the quantum number for the total angular
momentum of an eigenstate and $M$ is its $z$-component.  In order to
compare results on a finite sphere with those for an infinite planar
system, we identify the wavevector $q$ on the plane with the quantity
$L/R$, where $R$ is the radius of the sphere.  This identification is
exact in the limit $R \rightarrow \infty$, but it is to some degree
arbitrary for any finite system.  (For example, the identification $q
\rightarrow \sqrt{L(L+1)}/R$ would be equally valid.)

In keeping with the convention of the preceding companion
paper\cite{Simon}, we choose units of length such that the radius of
the sphere is unity. The effective ``size'' of the sphere is then
defined by the magnetic length $l_0 = S^{-1/2}$ where $S=N_{\phi}/2$
is half the total number of flux quanta through the surface of the
sphere.  It is also convenient to choose the unit of energy to be
$e^2/\epsilon l_0$ which is the only energy scale in the problem when
all of the electrons are confined to the lowest Landau level.

\subsection{Exact Diagonalization}
\label{sub:exact}

Since exact diagonalization in the spherical geometry is a well
established method, we will only describe it very briefly; more
information can be found in Refs. \cite{Haldane} and \cite{Fano}.  The
extraction of the projected dynamical structure factor
$\bar{S}(q,\omega)$ from such diagonalizations, however, is a method
new to this work.

As mentioned in the preceding companion paper\cite{Simon}, the single
particle energy eigenstates are the monopole spherical harmonics
\cite{Yang} $Y^{S}_{l,m}(\vO)$, where $l=S,S+1, \ldots,$ and
$m=-l,\ldots,l$. The corresponding eigenenergies are given by
\begin{equation}
E_l=\frac{1}{2S}[l(l+1)-S^2]\hbar\omega_c.
\end{equation}
where $\omega_c = \hbar S/m_b$ is the cyclotron frequency, with $m_b$ the
band mass of the electron.  The lowest Landau level has a basis of
the $2S+1$ states with $l=S$.
With the single particle states given,
ignoring the constant kinetic energy, the Hamiltonian of the
many electron system, projected to the lowest Landau level, is written
in second quantized notation as
\begin{equation}
\hat{H}=\frac{1}{2}\sum V_{m'_1,m'_2;m_1,m_2} C^{\dagger}_{S,m'_1}
C^{\dagger}_{S,m'_2}C_{S,m_2}C_{S,m_1},
\end{equation}
where $C^{\dagger}_{l,m}$ or $C_{l,m}$ creates or annihilates an electron in
the single particle eigenstate $Y^{S}_{l,m}(\vO)$, and $V_{m'_1,m'_2;m_1,m_2}$
is the matrix element of the spherical Coulomb interaction
$V(\vO_1,\vO_2)=\frac{1}{|\vO_1-\vO_2|}$,
namely
\begin{equation}
V_{m'_1,m'_2;m_1,m_2}=\int d\vO_1 d\vO_2
Y^S_{S,m'_1}(\vO_1)^{*}
Y^S_{S,m'_2}(\vO_2)^{*}
\frac{1}{|\vO_1-\vO_2|}
Y^S_{S,m_2}(\vO_2) Y^S_{S,m_1}(\vO_1),
\end{equation}
where the vectors $\vO$ are restricted to the surface of the sphere.
The relevant matrix elements are calculated more explicitly in Ref.
\cite{Fano}.

Now let us say a few words about the density operator in the spherical
geometry. By definition,
\begin{equation}
\rho(\vO)=\sum_i \delta(\vO-\vO_i),
\end{equation}
where $\vO_i$ is the position of the $i^{th}$ electron.  Since
$\rho(\vO)$ creates a neutral excitation when it acts on a state, it
is useful to expand it in {\em normal} spherical harmonics, {\it i.e.} we
define its Fourier components by
\begin{equation}
\rho_{lm}(\vO)=\sum_i Y_{l,m}(\vO_i),
\end{equation}
where $Y_{l,m}$ are the usual spherical harmonic (which are related to
the monopole spherical harmonics\cite{Yang} by $Y_{lm} = Y^0_{lm}$).
It is easy to verify that $\rho_{lm}$ has the expected properties:
\begin{equation}
\label{LL}
[L_z,\rho_{lm}]=m \rho_{lm},~~~
[L_{\pm},\rho_{lm}]=\sqrt{l(l+1)-m(m\pm 1)}~\rho_{lm\pm 1},
\end{equation}
and if $L^2 |\Phi\rangle=0$, then
\begin{equation}
L^2 \rho_{lm}|\Phi\rangle=l(l+1)\rho_{lm}|\Phi\rangle, ~~{\rm and}~L_z
\rho_{lm}|\Phi\rangle=m\rho_{lm}|\Phi\rangle.  \end{equation}

In second quantized notation, we can write the density operator as
\begin{equation}
   \rho(\vO) = \psi^{\dagger}(\vO) \psi(\vO)
\end{equation}
where $\psi^{\dagger}(\vO)$ and $\psi(\vO)$ are the operators that
create and annihilate respectively an electron at the point $\vO$.  As
usual, the eigenstate operators ($C^{\dagger}_{l,m}$)
can be easily related to the position space operators
($\psi^{\dagger}(\vO)$) via
\begin{equation}
  C^{\dagger}_{l,m} = \int d\vO \, Y^{S}_{lm}(\vO) \, \psi^{\dagger}(\vO),
\end{equation}
which can easily be inverted using the orthogonality property of the
monopole spherical harmonics\cite{Yang} to yield
\begin{equation}
  \psi^{\dagger}(\vO) = \sum_{l,m} [Y^S_{l,m}(\vO)]^*  C^{\dagger}_{l,m}.
\end{equation}
Thus we can write the density operator as
\begin{equation}
  \rho(\vO) = \sum_{l_1,m_1} \sum_{l_2,m_2} [Y^{S}_{l_1,m_1}(\vO)]^*
Y^S_{l_2,m_2}(\vO) C^{\dagger}_{l_1,m_1} C_{l_2,m_2}.
\end{equation}
We can then take the angular momentum components to yield
\begin{equation}
  \rho_{l,m} = \int d\vO \, Y_{l,m}(\vO) \, \rho(\vO) = \sum_{l_1,m_1}
\sum_{l_2,m_2}  \rho(l,m,l_1,m_1,l_2,m_2) C^{\dagger}_{l_1,m_1} C_{l_2,m_2}
\end{equation}
where
\begin{equation}
 \rho(l,m,l_1,m_1,l_2,m_2) = \int d\vO \, Y_{l,m}(\vO)
[Y^{S}_{l_1,m_1}(\vO)]^*  Y^S_{l_2,m_2}(\vO)
\end{equation}
which is evaluated explicitly in Appendix D of Ref \cite{Simon}.  It
is then easy to project this operator to the lowest Landau level ($l_1
= l_2 = S$) to get
\begin{equation}
  \label{eq:projden}
\bar{\rho}_{lm}=\sum_{m'} \bar{\rho}(l,m,m') C^{\dagger}_{S,m+m'}C_{S,m'},
\end{equation}
where
\begin{equation}
\label{eq:threej}
\bar{\rho}(l,m,m')=(-1)^{S+l+m+m'}(2S+1)
\sqrt{\frac{2l+1}{4\pi}}\tj{S}{S}{l}{-(m+m')}{m'}{m}\tj{S}{S}{l}{-S}{S}{0}.
\end{equation}

An important property of $\bar \rho_{lm}$ at $l=1$ is that
\begin{equation}
\label{LLL}
\bar{\rho}_{1,0}\propto \bar{L}_{z},~~\bar{\rho}_{1,1}\propto \bar{L}_{+},
{}~~{\rm
  and}~\bar{\rho}_{1,-1} \propto \bar{L}_{-},
\end{equation}
where the bar over the angular momentum operators indicate that they
have been projected to the first Landau level also.  This can be shown
by writing the angular momentum operators in terms of eigenstate
creation and annihilation operators\cite{Yang}, projecting to the
first Landau level and comparing to the definition of the projected
density operator.  For example, we can write
\begin{equation}
  \bar{L}_{z} = \sum_m  m C^{\dagger}_{S,m}C_{S,m}
\end{equation}
which is easily shown to be proportional to $\bar \rho_{1,0}$ once we have
evaluated the $3$-$j$ coefficient in Eq. [\ref{eq:threej}].  Using these
properties (Eqs. [\ref{LLL}]) of $\bar{\rho}_{1,m}$ we see that for
$|\Phi_0 \rangle$ in the first Landau level, $L^2 |\Phi_0\rangle = 0$
implies $\bar{\rho}_{1,m} |\Phi_0\rangle = 0$; {\it ie} the projected
density operator cannot generate excited states at $l=1$ from a
uniform state.

Now we have come to the central quantity of our calculation. By
definition, the projected dynamical structure factor is given by
\begin{equation}
\bar{S}(l,\omega)=
\sum_{\alpha}|\langle\Phi_{\alpha}|\bar{\rho}_{lm}|
\Phi_0\rangle|^2
\delta(\omega+\epsilon_0- \epsilon_{\alpha}),
\end{equation}
which is independent of $m$ by rotational invariance.  By
comparison, the full dynamical structure factor on a sphere is defined
by
\begin{equation}
S(l,\omega)=\sum_{\alpha}|\langle\Phi_{\alpha}|\rho_{lm}|\Phi_0\rangle|^2
\delta(\omega+\epsilon_0- \epsilon_{\alpha}).
\end{equation}
In the limit where the band mass is taken to zero, while the frequency
$\omega$ and the electron-electron interaction are held fixed, the
only energy states with finite $\epsilon_\alpha - \epsilon_0$ are
states in which all electrons are restricted to the lowest Landau
level.  Thus if the frequency $\omega$ is fixed on the scale of the
electron-electron interaction, the quantities $S(l,\omega)$ and $\bar
S(l,\omega)$ become identical, as mentioned earlier.  However, at
energies on the scale of the cyclotron frequency, transitions between
Landau levels will always be important, and the projected calculation
will be inaccurate.  In particular the projected structure factor will
never display a Kohn mode and will never satisfy the $f$-sum rule.

We should also mention that for the exact diagonalization it is
possible to take advantage of the rotational symmetry of the spherical
system to reduce the dimension of the relevant many-body Hilbert
space. Specifically, we explicitly block-diagonalize the Hamiltonian
by computing its matrix elements between many-body states with the
same total angular momentum $L^2$ and $L_z$. To construct an
eigenstate of the total angular momentum operator $L^2$ with
eigenvalue $l(l+1)$, we may in principle start from a randomly
generated many-body state with the required $L_z$ and use the operator
$\prod_{l'\neq l}[L^2-l'(l'+1)]$ to project out the unwanted
components with $L^2$ other than $l(l+1)$.  However, on a finite
precision computer, a naive application of this method is highly
unstable, {\it i.e.} the error introduced by the machine round-off
grows exponentially. Fortunately, there are ways to stabilize the
algorithm so that eigenstates of $L^2$ with up to machine precision
can be obtained. Details will be published elsewhere\cite{SONG}.

\subsection{Modified RPA}

\label{sub:ModRPA}

As mentioned in the preceding companion paper\cite{Simon}, there are
in principle two free parameters in the modified RPA. They are the
ratio of the effective mass to the band mass $m^*/m_b$, and a constant
$C=(\frac{\hbar^2}{m^*l_0^2})/(\frac{e^2} {\epsilon l_0})$, which is
the ratio of the effective quasiparticle cyclotron energy to the
Coulomb energy (which should be of order one).  The estimate $C
\approx 0.3$ has been made by Halperin, Lee, and Read\cite{HLR} by
examining results from exact diagonalizations of small systems.  We
will use this value of $C$ for all of our calculations in this paper.

In order to project our results to the lowest Landau level, we should
take the band mass to zero, or $m^*/m_b \rightarrow \infty$.  In
practice, this is calculationally difficult, and we have actually used
$m^*/m = 50$ which is quite sufficient to separate the Kohn mode from
the low energy excitations.  We then use the prescription described
the preceding companion paper\cite{Simon} to calculate the
density-density response function $K_{00}(L,\omega)$ which is related
to the dynamical structure factor via
\begin{equation}
  \label{eq:relate}
  S(L,\omega) = \frac{-1}{\pi} \mbox{Im}[K_{00}(L,\omega)].
\end{equation}
  As noted in \cite{Simon}, the modified
RPA is constructed to satisfy the $f$-sum rule, which one derives from
the exact structure factor on the sphere.  Within the units employed
in this paper, this becomes
\begin{equation}
\label{eq:fsum}
  \int_0^{\infty} d\omega\ \omega S(L,\omega)=\frac{L(L+1)}{8 \pi}\frac{N
m^*}{m_b}  \frac{C}{S},
\end{equation}
where the factor $\frac{C}{S}$ (with $S = \frac{N_{\phi}}{2}$ here) is
simply a conversion of energy scales as described above.  Note that
this quantity diverges in the limit $m_b/m^* \rightarrow 0$.  The main
contribution to this sum rule, however, is from the Kohn mode at (or
above) the cyclotron frequency.  All other modes described by the
modified RPA are low energy excitations (on the energy scale set by
the Coulomb interaction).  Moreover, as discussed above, in the exact
diagonalizations, the projection to the lowest Landau level restricts
our results to only those modes on the energy scale of the Coulomb
interaction.  Thus, in order to ``project'' $S(L,\omega)$ calculated
in the modified RPA to the first Landau level, we simply discard the
high frequency Kohn mode and call the result the projected dynamical
structure factor $\bar S(L,\omega)$.  We then define a projected
contribution to the $f$-sum rule by
\begin{equation}
\label{fsum}
\bar f(L)=\int_0^{\infty} d\omega\ \omega \bar{S}(L,\omega).
\end{equation}
The value of $\bar f(L)$ remains finite for $m_b/m^* \rightarrow 0$
for both the exact calculation and the modified RPA, so it is
reasonable to try to compare the two.

In a similar manner, we define the projected static structure factor
$\bar S(L)$, and the contribution $\Lambda(L)$ to the compressibility
sum rule by
\begin{eqnarray}
\bar{S}(L)=\int_0^{\infty} d\omega\ \bar{S}(L,\omega),
\label{eq:SL} \\
\label{csum}
\Lambda(L)=\int_0^{\infty} d\omega \frac{\bar{S}(L,\omega)}{\omega}.
\end{eqnarray}
In addition we shall report results for the average frequency entering
in the single mode approximation (SMA)\cite{Girvin}, which is simply
defined by
\begin{equation}
  \label{eq:SMA}
  \omega_L=\frac{\int_0^{\infty} d\omega\ \omega
\bar{S}(L,\omega)}{\int_0^{\infty} d\omega
\bar{S}(L,\omega)} = \frac{\bar f(L)}{\bar S(L)}.
\end{equation}

The complete static structure factor $S(L)$ is defined using
$S(L,\omega)$ on the right hand side of Eq. (\ref{eq:SL}) instead of
$\bar S(L,\omega)$.  The contribution to $S(L)$ from the Kohn mode, or
from inter-Landau-level transitions does not diverge in the limit
$m_b/m^* \rightarrow 0$, but it is a large finite contribution;
therefore, to make sensible comparisons between the modified RPA and
the exact calculations for the lowest Landau level, we use the
projected structure factor in both cases.

By contrast, in the limit $m_b/m^* \rightarrow 0$, the contribution to
$\Lambda(L)$ from the Kohn mode or from inter-Landau-level transitions
is vanishingly small.  Therefore, it would not matter whether we use
$\bar S(L,\omega)$ or $S(L,\omega)$ on the right hand side of Eq.
(\ref{csum}).

\section{Numerical Results}
\label{Results}

In this section we present and compare numerical results obtained from
exact diagonalizations and from the modified RPA, for two filling
factors in the principal sequence $\nu_p=\frac{p}{2p+1}$. The first
set of data, shown in Figs. \ref{F1a}-\ref{F1e}, is obtained for a
system of eight electrons at a filling factor $\nu=1/3$ (As required
by Eqs.  [\ref{eq:Nshift1}] and [\ref{eq:Nshift2}], $N_{\phi} =21$
here).  In the composite fermion picture, at the mean field level, the
groundstate of the system corresponds to a completely filled lowest
effective Landau level for the composite fermions. The second set of
data, shown in Figs. \ref{F2a}-\ref{F2e}, is obtained for a system of
twelve electrons at a filling factor $\nu=3/7$ ($N_{\phi} = 23$),
where the groundstate of the system corresponds to the lowest three
filled effective Landau levels of the composite fermions.

As discussed in the Refs. \cite{Simon,HLR}, and \cite{Simon2}, in a
planar system, we expect the modified RPA to be more accurate in the
large $p$ limit (ie, $\nu$ approaching $1/2$) where the motion of the
composite fermions is semiclassical.  In the present calculations
however, we do not necessarily expect the modified RPA to be more
accurate for the $\nu=3/7$ ($p=3$) state than for the $\nu=1/3$
($p=1$) state due to the problem of enhanced finite size effects for
the $\nu=3/7$ state.  Although the $\nu = 3/7$ system contains more
electrons ($N=12$) than the $\nu = 1/3$ system ($N=8$), it should be
considered the ``smaller'' system in that the composite fermion
magnetic length is much larger in the $\nu = 3/7$ system, and the
number of electrons per Landau level is also smaller.  Thus, any
finite size effects will be accentuated in the $\nu=3/7$ set of data,
and we suspect that the modified RPA will be less accurate in this
case.

In Fig. \ref{F1a}, we show the projected spectral weights of
$\bar{S}(L,\omega)$ for a system of eight electrons at a filling
factor $\nu=1/3$ ($N_{\phi}=21$), calculated through exact
diagonalization (upper graph) and using the modified RPA (lower
graph). The weight of each delta function contribution to
$\bar{S}(L,\omega)$ at each $(L,\omega)$ is proportional to the area
of the corresponding shaded rectangle.  Here and in the rest of this
work, the energy scale is given in units of $e^2/(\epsilon l_0^2)$.
As we discussed above, although there may be excitations at $L=1$,
these states all  have zero weight in the projected dynamical
structure factor.

We see immediately from Fig. \ref{F1a} that the projected dynamical
structure factor calculated in the modified RPA agrees very well with
the exact diagonalization at low energies.  In particular, the
distribution of spectral weights among the states in the lowest
collective mode is very similar in the upper and lower graph.  The
small discrepancy in overall energy scales might be due to an
inaccuracy in our guess of the energy scale conversion factor $C$
described above.

We note that the lowest collective mode dominates the spectral weight
for angular momenta up to $L=L_{roton} \approx 4$ in the exact
diagonalization, particularly near the roton minimum $L_{roton}$.  In
the modified RPA the lowest mode dominates only at smaller angular
momenta.  At higher angular momenta ($L > L_{roton}$), the exact
diagonalization shows a fairly significant amount of spectral weight
at energies $E\sim 2E_{roton}$. It is believed that the corresponding
states can roughly be represented as composite objects of two rotons
\cite{SH}.  We should not expect that such states would be properly
modeled in the modified RPA since the RPA only represents single
quasiparticle-quasihole excitations.  However, as long as we consider
sufficiently low energy (not too much larger than $E\sim 2E_{roton}$),
the modified RPA seems to agree very well with the exact
diagonalization.

At higher energies, in the continuum of the exact diagonalization
spectra, we empirically observe that the spectral weights decay
exponentially $\bar{S}(L,\omega)\sim e^{-\omega/\Gamma}$ for a given
$L$, where $\Gamma \sim 0.03e^2/\epsilon l_0$.  In this high energy
region, where the modified RPA differs significantly from the exact
results, we see that the modified RPA yields a discreet spectrum of
high energy modes with a large amount of spectral weight.  The
discreteness of the modified RPA is an obvious artifact arising from
the neglect of processes where a higher energy mode can decay into
several low energy modes with the same total energy and wavevector.
The fact that the modified RPA seriously overestimates the total
weight at high energies is a more significant limitation of the
approximation.

In Fig. \ref{F1b} we show unweighted excitation spectra for the same
system of $N=8$ electrons at $\nu=1/3$. The upper graph is the
complete energy spectrum in the range $0 \le \omega < 0.45$, $0 \le L
< 11$, from exact diagonalization regardless of whether or not the
state contributes to $\bar S(q,\omega)$. The lower graph is again the
result of the modified RPA.  Note that the dispersion relations and
the overall energy scale of the lowest collective mode are very
similar in the two cases.  The black circles in the two graphs are the
spectra of the single mode approximation (SMA) computed from with Eq.
[\ref{eq:SMA}], for the exact calculations and modified RPA
respectively.  We observe that in the exact diagonalization results,
the SMA works well for angular momenta up to the roton minimum
$L_{roton}$ while in the modified RPA, the SMA fails at all but the
lowest angular momentum.  This failure of the SMA is due to the erroneous
predictions of the modified RPA at high energy.  In the modified RPA,
as we increase the angular momentum, the main contribution to the
spectral weight comes from a frequency that increases approximately
proportionally to $L(L+1)$.  This can be explained roughly by
considering the modified RPA as a perturbation of a free electron
system where it is well known that the excitations are confined to a
band whose upper and lower energy boundaries both vary as $q^2$.  When
a magnetic field is added to such a free electron system, a Kohn mode
appears, and some of the low energy excitations can be changed, but we
expect that most of the high $q$ excitations remains roughly in
the same band.  As discussed above, the single particle nature of the
modified RPA allows these high energy modes to be given too much
spectral weight.  Thus we see how the the single mode approximation in
the modified RPA is dominated by the poorly modeled higher frequencies
in the modified RPA, thus giving very inaccurate results.

In Fig. \ref{F1c}, we show the projected static structure factor
$\bar{S}(L)$ (see Eq. [\ref{eq:SL}]) for the same system of $N=8$
electrons at $\nu=1/3$.  The circles (solid line) are from the exact
diagonalization whereas the triangles (dotted line) are the result of
the modified RPA.  The inset is the same modified RPA data on a
reduced scale to more clearly show the behavior at higher angular
momenta.  Of course the projected static structure factor is only
defined at integer values of $L$, and the lines (dotted and solid) are
just guides for the eye.  In the exact diagonalization (circles), we
find that $\bar{S}(L)$ peaks around the roton minimum and decreases
rapidly at higher angular momenta.  We have studied its behavior and
found empirically that it roughly obeys a Gaussian $\bar{S}(L)\sim
e^{-\alpha L(L+1)}$, where $\alpha$ is a constant.  This reflects a
similar empirically observed Gaussian decay of the matrix elements of
the density operator at large angular momentum.  The projected static
structure factor in the modified RPA (triangles) is very similar to
the exact result (circles) at small values of the angular momentum
where the modified RPA does not predict extraneous high energy modes
with large amounts of weight.  At higher values of $L$ the erroneous
high energy modes cause a severe overestimation of the projected
static structure factor.

In Figs. \ref{F1d} and \ref{F1e}, we show the contribution to the
$f$-sum rule and compressibility sum rule as defined in Eqs.
[\ref{fsum}] and [\ref{csum}] for the same system of $N=8$ electrons
at filling fraction $\nu=1/3$.  As above, the circles are from the
exact diagonalization whereas the triangles are the result of the
modified RPA, and the lines are guides for the eye.  Once again the
insets show the same modified RPA data on a reduced scale.  In Fig.
\ref{F1d}, we see that the contribution to the $f$-sum rule suffers
from the same erroneous high energy modes.  The compressibility sum
rule (Fig. \ref{F1e}) is somewhat better estimated by the modified RPA
because the contribution of the erroneous high energy modes are
suppressed by a factor of $\omega$ in this sum rule (see Eq.
[\ref{csum}]).  This is somewhat encouraging, since the
compressibility sum is the most relevant of these quantities for
describing the low energy behavior of the quantized Hall state.

In Fig. \ref{F2a}, we show the projected dynamical structure factor
$\bar{S}(L,\omega)$ for a system of $N=12$ electrons at a filling
factor $\nu=3/7$ ($N_{\phi} = 23$). Again, the spectral weights of
$\bar{S}(L,\omega)$ at each $(L,\omega)$ are proportional to the area
of the corresponding shaded rectangles.  We observe again that the
behaviors of $\bar{S}(L,\omega)$ from exact diagonalization (upper
graph) and calculated in the modified RPA (lower graph) are
semi-quantitatively similar.  However, once again, the modified RPA
results clearly overestimate the spectral weights at high energies.

In Fig. \ref{F2b}, we show the corresponding complete excitation
spectra of the system of $N=12$ electrons at a filling factor
$\nu=3/7$.  An interesting feature to note is that there are wiggles
in the dispersion of the lowest collective mode in both the exact
calculation and the modified RPA. Note particularly how the lowest
energies at $L = 3$ and $5$ lie above the lowest energies at $L = 2,
4,$ and $6$.  This behavior was observed previously in the exact
calculation for this system size and filling factor by d'Ambrumenil
and Morf\cite{Morf}.  These authors posed the question whether the
even-odd alternation might be a spurious effect due to the finite size
of the system.  Based on the modified RPA analysis, however, we
believe that the maxima and minima in the spectrum are genuine effects
reflecting the maxima and minima that were previously
found\cite{Simon2} in the dispersion relation for the lowest
excitation branch in the planar system using this approximation.  More
generally, for filling fractions of the form $\nu = p/(2p+1)$, at
large values of $p$, it was predicted in Ref. \cite{Simon2} that there
should be a series of maxima and minima in the dispersion of the
lowest excitation branch occuring respectively at wavevectors of the
form $q = \pi (n - \frac{1}{4})/ R_c^*$ and $q = \pi (n +
\frac{1}{4})/ R_c^*$, where $n$ is a positive integer, and $R_c^*$ is
the ``effective cyclotron radius'' for the composite fermions.  This
quantity is given by the relation
\begin{equation}
  R_c^* = \hbar k_{\mbox{\tiny F}}/(e \Delta B) = 2p/k_{\mbox{\tiny
        F}},
\end{equation}
where $\Delta B$ is the deviation of the magnetic field from the field
at $\nu = 1/2$, and the Fermi wavevector $k_{\mbox{\tiny F}}$ is
related to the electron density $n_e$ by $k_{\mbox{\tiny F}} = (4 \pi
n_e)^{1/2}$.  Thus, at filling fraction $\nu = p/(2p+1)$, we have
\begin{equation}
  R_c^* = \frac{1}{l_0} (2p)^{1/2} (2p+1)^{1/2}.
\end{equation}
If we now apply this formula to our system of particles on a sphere,
with $l_0^{-1} = (N_{\phi}/2)^{\frac{1}{2}}$, we find $R_c^* \approx
2$ at $\nu = 3/7$, with $N=12$. Thus one might expect maxima and
minima to alternate with a period of $\Delta L = \pi/ R_c^* \approx
1.6$.  Since the observed alternation with a period of $\Delta L = 2$
is not far from this expectation, we believe it is appropriate to
identify the oscillations as the expression in our finite size system
of the oscillations predicted for the plane.  In any case, the fact
that this detail of the spectrum predicted from the modified RPA for
the finite system is in good agreement with the exact calculations
gives support to the belief that the oscillations predicted for the
infinite system should also be present in an exact calculation.

The black circles in Fig. \ref{F2b} show the single mode approximation
(Eq. [\ref{eq:SMA}]) for $\nu = 3/7$. As we discussed in the case of
$\nu=1/3$, the single mode approximation clearly shows that the
modified RPA predicts too much spectral weight at high energies.

In Figs. \ref{F2c}, \ref{F2d}, and \ref{F2e}, we show the projected
static structure factor (Eq. [\ref{eq:SL}]), the contribution to the
$f$-sum rule (Eq. [\ref{fsum}]) and the contribution to the
compressibility sum rule (Eq. [\ref{csum}]) for the same system of
$N=12$ electrons at filling fraction $\nu=3/7$.  As above, the circles
are always from the exact diagonalization and the triangles are from
the modified RPA.  Once again the insets show the modified RPA data on
a reduced scale to show the behavior at higher angular
momenta.  Again, we see that the erroneous high energy modes in the
modified RPA dominate the $f$-sum and the static structure factor at
all but the lowest angular momenta.  On the other hand, the
compressibility sum (Fig. \ref{F2e}) is somewhat better represented in the
modified RPA due to the suppression of these high energy modes.

\section{Conclusions}
\label{C}

We have studied the excitation spectra and the dynamical structure
factor in quantum Hall states belonging to the principal sequence $\nu
= p/(2p+1)$ through exact diagonalization and projected modified RPA.
In particular, we have presented results from finite size exact
diagonalization and from projected modified RPA on a system of $N=8$
electrons at $\nu=1/3$ and a system of $N=12$ electrons at $\nu=3/7$.
We find that the modified RPA works reasonably well at low energies,
ie for energies $\lesssim 0.2$, in units of $e^2/\epsilon l_0$. (By
comparison, the effective Fermi energy $E_{\mbox{\tiny F}}=
k_{\mbox{\tiny F}}^2/(2m^*)$ is approximately $C/2 \approx 0.15$, in
these units.)

By combining our results with previous analyses, we can arrive at some
reasonable conjectures for the excitation spectrum in a planar system,
which should hold in particular for relatively large values of the
parameter $p$.

1.   The present work supports the conjecture that the modified RPA
gives a qualitatively correct description of the dispersion curve for
the lowest branch of the excitation spectrum (quasiexciton mode) at
$\nu = p/(2p+1)$.  The series of maxima and minima predicted by the
modified RPA in the planar limit have analogs in the finite system
which are also observed in the exact calculations.

2.  The general distribution of the spectral weight $S(q,\omega)$ for
frequencies above the lowest excitation branch, up to energies of the
order of $\omega \approx 0.2 e^2/(\epsilon l_0)$, for wavevectors in
the range $0 \le q \le 2 k_{\mbox{\tiny F}}$ is likely to be
represented in a qualitatively correct fashion by the modified RPA.
Of course certain details must clearly be wrong as has been previously
discussed\cite{Simon2}.  Higher branches of the energy spectrum, which
are undamped in the modified RPA, should actually be broadened into a
continuum by decay into multiple excitations of lower lying modes.

3.  The modified RPA predicts that for large values of $p$, and
wavevectors in the range $k_{\mbox{\tiny F}}/p \ll q \ll
k_{\mbox{\tiny F}}$, the dominant contributions to the static
compressibility sum rule should arise from frequencies $\omega \propto
q^2$. \cite{HLR}  We believe this prediction to be correct, but our
finite size systems are too small to give any direct confirmation.

4.  The modified RPA predicts that there should be sizable
contributions to $\bar S(q,\omega)$, for $\omega \gtrsim 0.2
e^2/(\epsilon l_0)$, for large wavevectors in the range where
$ | q^2 - 2 m^* \omega | \lesssim 2 k_{\mbox{\tiny F}} q$.  We believe
these predictions to be spurious, however, as exact finite size system
calculations show little weight in $\bar S(q,\omega)$ at high
frequencies for any value of $q$.  Of course the complete dynamical
structure factor $S(q,\omega)$ will contain additional contributions
at very high frequencies arising from transitions between Landau levels.

5.  The above conjectures are of course predicated on the assumption
that no instability or phase transition occurs at very large  values
of $p$, which would then invalidate our analysis.  Finite size system
calculations cannot rule out the possibility of such an instability at
large $p$.

\vspace{10pt}

{\centerline{\small ACKNOWLEDGMENTS}}

The work at Harvard was supported by the National Science Foundation
Grant DMR-91-15491.

\begin{figure}~\caption{}~{The weights of the projected dynamical structure
factor
    $\bar{S}(L,\omega)$ for $N=8$ electrons at filling fraction
    $\nu=1/3$ ($N_{\phi}=21$).  The upper graph is obtained from
    exact diagonalization, and the lower graph is obtained from
    modified RPA.  The magnitude of each delta function contribution
    to the projected dynamical structure factor $\bar{S}(L,\omega)$ is
    proportional to the area of the corresponding shaded rectangle.  }
\label{F1a}
\end{figure}

\begin{figure}~\caption{}~{The excitation spectra for $N=8$ electrons at
filling fraction
    $\nu=1/3$. The upper graph is obtained from finite size exact
    diagonalization, and the lower graph is obtained from modified
    RPA.  The black circles are spectra calculated in the single mode
    approximation (see Eq. [\ref{eq:SMA}]).}
\label{F1b}
\end{figure}

\begin{figure}~\caption{}~{The projected static structure factor $\bar{S}(L)$
(See Eq. [\ref{eq:SL}]) for
  $N=8$ electrons at filling fraction $\nu=1/3$.  The circles are
  obtained from finite size exact diagonalization, and the triangles
  are obtained from the modified RPA.  The inset is the modified RPA
  data on an expanded scale to show the behavior at large $L$.  The
  solid and dotted lines are guides for the eye.}
\label{F1c}
\end{figure}

\begin{figure}~\caption{}~{The contribution to the $f$-sum rule $\bar f(L)$
(See Eq. [\ref{fsum}]) for
    $N=8$ electrons at filling fraction $\nu=1/3$.  The circles are
    obtained from finite size exact diagonalization, and the triangles
    are obtained from the modified RPA.  The inset is the modified RPA
    data on an expanded scale to show the behavior at large $L$.  The
    solid and dotted lines are guides for the eye.}
\label{F1d}
\end{figure}

\begin{figure}~\caption{}~{The contribution to the compressibility sum rule
$\Lambda(L)$ (See
  Eq.  [\ref{csum}]) for $N=8$ electrons at filling fraction $\nu=1/3$.
  The circles are obtained from finite size exact diagonalization, and
  the triangles are obtained from the modified RPA.  The inset is the
  modified RPA data on an expanded scale to show the behavior at large
  $L$.  The solid and dotted lines are guides for the eye.}
\label{F1e}
\end{figure}

\begin{figure}~\caption{}~{ The projected dynamical structure factor
$\bar{S}(L,\omega)$ for
  $N=12$ electrons at filling fraction $\nu=3/7$ ($N_{\phi} = 23$).
  The upper graph is obtained from exact diagonalization, and the
  lower graph is obtained from modified RPA.  The magnitude of each
  delta function contribution to the projected dynamical structure
  factor $\bar{S}(L,\omega)$ is proportional to the area of the
  corresponding shaded rectangle.}
\label{F2a}
\end{figure}

\begin{figure}~\caption{}~{ The excitation spectra for $N=12$ electrons at
filling fraction
  $\nu=3/7$. The upper graph is obtained from finite
  size exact diagonalization, and the lower graph is obtained from
  modified RPA.  The black circles are spectra calculated in the
  single mode approximation (see Eq. [\ref{eq:SMA}])}
\label{F2b}
\end{figure}

\begin{figure}~\caption{}~{The projected static structure factor $\bar{S}(L)$
(See Eq. [\ref{eq:SL}]) for
  $N=12$ electrons at filling fraction $\nu=3/7$.  The circles are
  obtained from finite size exact diagonalization, and the triangles
  are obtained from the modified RPA.  The inset is the modified RPA
  data on an expanded scale to show the behavior at large $L$.  The
  solid and dotted lines are guides for the eye.}
\label{F2c}
\end{figure}

\begin{figure}~\caption{}~{The contribution to the $f$-sum rule $\bar f(L)$
(See Eq. [\ref{fsum}]) for
  $N=12$ electrons at filling fraction $\nu=3/7$.  The circles are
  obtained from finite size exact diagonalization, and the triangles
  are obtained from the modified RPA.  The inset is the modified RPA
  data on an expanded scale to show the behavior at large $L$.  The
  solid and dotted lines are guides for the eye.}
\label{F2d}
\end{figure}

\begin{figure}~\caption{}~{The contribution to the compressibility sum rule
$\Lambda(L)$ (See
    Eq.  [\ref{csum}]) for $N=12$ electrons at filling fraction
    $\nu=3/7$.  The circles are obtained from finite size exact
    diagonalization, and the triangles are obtained from the modified
    RPA.  The inset is the modified RPA data on an expanded scale to
    show the behavior at large $L$.  The solid and dotted lines are
    guides for the eye.}
\label{F2e}
\end{figure}

\end{document}